\documentclass[11pt,a4paper]{JHEP}
\usepackage{amssymb} 

\def\bX{\overline X}
\def\l{\lambda}

\def\d{\partial}

\def\bX{\overline X}
\def\l{\lambda}

\newcommand\0{\nonumber}
\newcommand\pp{{\mathbb P}}
\newcommand\rr{{\mathbb R}}
 
\newcommand\zz{{\mathbb Z}}
\newcommand\cc{{\mathbb C}}
\newcommand\ee{\end{eqnarray}}	 	
\newcommand\be{\begin{eqnarray}}
\newcommand\ba{\begin{array}}			
\newcommand\ea{\end{array}}
\newcommand\eeq{\end{equation}}	 	
\newcommand\beq{\begin{equation}}

\preprint{SISSA/155/99/EP/FM\\SPIN-1999/32\\\tt hep-th/9912227}

\title{Instantons and scattering in ${\cal N}=4$ SYM in 4D} 

\author{G.Bonelli\\
Spinoza Institute, University of Utrecht,
Leuvenlaan 4, 3584, CE Utrecht, The Netherlands\\
E-mail:  \email{G.Bonelli@phys.uu.nl}}

\author{ L.Bonora, S.Terna, A.Tomasiello\\
International School for Advanced Studies (SISSA/ISAS)\\
Via Beirut 2--4, 34014 Trieste, Italy, and INFN, Sezione di Trieste\\
E-mail:   \email{bonora@sissa.it}, \email{terna@sissa.it}, 
\email{tomasiel@sissa.it} }

\abstract{We study classical solutions (ic--instantons) in ${\cal N}=4$ 
SYM in 4D which, in the strong coupling limit, correspond to complex 
two--dimensional manifolds. Asymptotically in time the latter have 
boundaries represented by compact real three--manifolds. Therefore they 
lend themselves to an interpretation in terms of 3--brane scattering. We 
suggest that these solutions may represent scattering of D3--branes of
type IIB theory in 10D. In particular we show that the world--volume theory on 
complex two--dimensional manifolds is the correct one for D3--branes.}


\begin{document}

\section{Introduction}

In this paper we discuss a new type of classical solutions in ${\cal N}=4$ SYM 
theory in 4D with $U(N)$ gauge group.
This theory is well--known for its self--duality properties \cite{MO}
and its duality properties with type IIB supergravity (superstring) theory
via AdS/CFT correspondence, \cite{malda}, are still under intensive study.
Here we show that there is a (still unexplored) nonperturbative sector of the
theory based on a new type of instantons. This is partly parallel to what 
happens in ${\cal N}=(8,8)$ 2D theory with gauge group $U(N)$, which has been 
called Matrix String Theory (MST), \cite{DVV}. In MST one finds classical 
solutions that, in the strong coupling limit, 
become Riemann surfaces with punctures, which are natural candidates 
to represent scatterings of closed strings, \cite{wynter,GHV,bbn1}. 
This idea was confirmed by the subsequent analysis in 
\cite{bbn2,bbnt1,bbnt2,wynter2}. This led to the identification of the strong 
coupling limit of MST with perturbative type IIA theory. The solutions 
in question were called {\it stringy} or {\it Riemannian instantons}.
 
Similar classical solutions  can be found in other dimensions. In this paper
we deal with 4D. In view of these generalizations the world {\it instanton}
may sound misleading, therefore we will use for these new kind of solutions 
the term {\it interaction--carrying instantons} or simply
{\it ic--instantons}. The reason we keep calling them generically instantons 
is due to the analogy with ordinary instantons: just as the latter are thought 
to represent interpolating solutions between different vacua, we think of 
ic--instantons as interpolating solutions between given initial and 
final asymptotic states.

In this paper we construct such ic--instantons in 4D SYM and we suggest that 
the ic--instantons of ${\cal N}=4$ SYM theory
with gauge group $U(N)$, in the strong YM coupling limit, may represent scattering
processes involving 3--branes, which we identify as D3--branes of type IIB 
theory in 10 dimensions. In support of this suggestion we show that 
ic--instantons at strong coupling describe branched 
coverings of the 4 dimensional base manifold which we assume to have a 
complex structure. These branched coverings are complex 2 dimensional 
surfaces\footnote{Throughout the paper, by surface without qualifier we mean 
a complex two--dimensional surface. Whenever we want to indicate a real 
two--dimensional surface we use the term 'Riemann surface'.}
with boundaries which have the correct geometry to describe scatterings 
of 3--branes. Moreover we show that the world--volume theory on the surface is
the correct one for D3--branes. Finally we show that the sum over ic--instantons
gives rise to a series weighted by powers of the inverse YM coupling constant,
with an exponent given by the Euler characteristics of the corresponding 
surfaces. The analysis of this last part is largely incomplete and what
we report in this paper can only be considered as a  preliminary exploration 
on this subject. 

The paper is organized as follows. In section 2 we introduce the notation and
derive in various ways the equations for ic--instantons. In section 3 we
describe some general properties for ic--instantons. Section 4 is devoted to
the explicit construction of such classical solutions; in particular we
illustrate the factorization of ic--instantons in a {\it group theoretical
factor} and a {\it branched covering} factor. In the strong coupling limit
only the second factor is relevant. In section 5 we discuss some general
properties and give a few explicit examples of branched coverings.
In section 6 we expand the SYM action
about an ic--instanton solution in the strong coupling limit: we find that
the dominant part of the action is lifted to the covering surface, say $\Sigma$,
and becomes the action of ${\cal N}=4$ free supersymmetric Maxwell theory. 
We show that the amplitude induced by an ic--instanton corresponding to
$\Sigma$ is proportional to a power of the inverse YM coupling constant 
whose exponent is the Euler characteristics of $\Sigma$. This result comes 
from a counting of zero modes over $\Sigma$. In section 7  we discuss the
relation with Matrix Theory and other problematic or unresolved questions.

\section{Interaction--carrying classical solutions}

The Minkowski action of ${\cal N}=4$ SYM theory in 4D is
\be
S&=& \int_{\cal X} d^4x~{\rm Tr}~\left( 
		-\frac{1}{4g^2}F_{\mu\nu} F^{\mu\nu}
                -\frac{1}{2}D_{\mu}X^i D^{\mu}X_i
		+\frac{g^2}{4}\left[X^i,X^j\right]^2 
		+\frac{i}{2} \bar{\l}\gamma^\mu D_\mu\l\right.\0\\
&&\left. +\frac{g}{4}\left(\l^T C{\gamma^i}^\dagger\left[X^i,\l\right]
    -{\l}^\dagger C\gamma^i\left[X^i,{\l}^*\right]\right)\right)\label{mSYM}
\ee
where $i=1,\ldots,6$. ${\cal X} $ is a four dimensional manifold of the type
${\cal X} = {\mathbb R}\times M_3$, where $M_3$ is a three--dimensional
compact manifold and ${\mathbb R}$ is the line $-\infty<x^0<\infty$. Although
the action (\ref{mSYM}) can be studied on more general manifolds, we will
consider in the following essentially two examples: $M_3=S^3$, the 3--sphere, 
and $M_3= {\mathbb T}^3$, the 3--torus defined by periodic $x^1,x^2,x^3$. 
We always suppose that ${\cal X}$ admit complex structures. 
 
$F^{\mu\nu}$ is the field strength of the gauge field $A_\mu$, the $X^i$ are 
$N\times N$ hermitean matrices in the adjoint of $U(N)$; from a geometrical 
point of view, we understand the existence of a vector bundle $E$, with structure
group $U(N)$, so that $X^i$ are sections of $End E$. $\l$ is an 
$N\times N$ matrix whose entries are both Weyl 
spinors of $SO(1,3)$ and vectors in the fundamental of 
$SU(4)$: namely the $\gamma^\mu$'s will act on the $SO(1,3)$ spinorial indices,
while the $\gamma^i$'s on the $SU(4)$ ones. Since we make explicit use of them
in the following, we write down our definitions for the gamma matrices:
\be
&&\gamma^\mu=\left(\matrix{0               & \sigma^\mu\cr
		  	 \bar{\sigma}^\mu& 0         \cr}\right)  ,  
\quad\quad\Gamma^i=\left(\matrix{0                   & \gamma^i\cr
		       {\gamma^i}^\dagger & 0       \cr}\right)\0
\ee
where $\sigma^0=-\bar{\sigma}^0=\mathbf{1}$, $\sigma^i=\bar{\sigma}^i$ are the 
Pauli matrices; the $\Gamma^i$ are the $8\times $8 6D gamma matrices as in 
\cite{sohnius} and $C$ is the 4D charge conjugation matrix; they satisfy 
the usual anticommutation relations:
\be
&&\left\{ \gamma^\mu,\gamma^\nu\right\}=2 \eta^{\mu\nu}  ,
\quad\quad  \left\{ \Gamma^i,\Gamma^j\right\}=2 \delta^{ij}.
\ee

The supersymmetric transformations are
\be
&&\delta X^i =\frac{i}{g}\left(\epsilon^T C{\gamma^i}^\dagger\l-
			\epsilon^\dagger C\gamma^i\l^*\right)   \0\\
&&\delta A_{\mu}=-i\left(\bar{\epsilon}\gamma^\mu\l-
			\bar{\l}\gamma^\mu\epsilon\right)   \0\\
&&\delta \l=-\frac{1}{g^2}F_{\mu\nu}\gamma^{\mu\nu}\epsilon
	    -i\left[X_i,X_j\right]\gamma^{ij}\epsilon
	    +\frac{2}{g}D_\mu X_i \gamma^\mu \gamma^0 C \gamma^i \epsilon^*.
	    \label{msusy}
\ee

Ic--instantons in 2D can appear either as BPS classical solutions of the SYM 
theory or as 4D self--dual systems reduced to 2 dimensions (by the way, this 
is another reason why such solutions were called instantons). Analogously, in 4D,
ic--instantons can be seen either as classical solutions that preserve
part of the supersymmetry or as 8D self--dual systems
reduced to 4D.  

In the next subsection we will discuss classical solutions that preserve some
supersymmetry. We can follow two courses: either we use the Minkowski 
supersymmetric transformations (\ref{msusy}) above, find classical equations 
whose solutions preserve a fraction of supersymmetry and Wick--rotate such 
equations to their Euclidean form; or we can use the Euclidean version 
of the supersymmetry transformations and find equations whose solutions preserve
some supersymmetry. For simplicity we take the second course but the result 
is the same in both cases. However passing to the Euclidean formulation 
introduces a well--known problem in supersymmetric theories. The Euclidean 
transcription of a supersymmetric Minkowski theory may considerably modify
the supersymmetric properties of the latter if Weyl/Majorana fermions are 
involved, which is the case here. There are several recipes to deal with this 
problem, see \cite{VN} and references therein. We will follow \cite{VN}:
such an approach amounts to an effective doubling of the degrees of freedom
of the Euclidean version with respect to the Minkowski one.

With some abuse of language we will call the above solutions BPS solutions,
in the sense of supersymmetry preserving solutions. This is substantially
motivated by the fact that the final theory on the covering space at
strong coupling will turn out to be supersymmetric (see below).

\subsection{Ic--instanton equations as BPS solutions}

We write first the Euclidean action in terms of the complex coordinates 
$v=\frac 12 (x^1+ix^2)$, $w= \frac 12 (x^3+ix^4)$,
\be
S&=&  \int_{\cal X} d^2vd^2w~{\rm Tr}~\left( D_vX^iD_{\bar v}
 X^i + 
D_wX^iD_{\bar w} X^i  -\frac{g^2}{2} [X^i,X^j]^2 \right.\0\\
&&- \frac{1}{4g^2} (F_{v\bar v}^2 + F_{w\bar w}^2 - 2 F_{vw}F_{\bar v 
\bar w} - 2 F_{v\bar w}F_{\bar v w})\0\\
&&-2\left(\l_1^* D_{\bar v}\l_1+\l_2^* D_v\l_2\right)
-2\left(\l_1^* D_{\bar w}\l_2-\l_2^* D_w\l_1\right) \0\\
&&\left.-\frac{g}{2}\left(\l^T C{\gamma^i}^\dagger\left[X^i,\l\right]
    -{\l}^\dagger C\gamma^i\left[X^i,{\l}^*\right]\right)\right) \label{eSYM}
\ee
Next we write the Euclidean version of the ${\cal N}=4$ supersymmetric
transformations
\be
&&\delta X^i =\frac{i}{g}\left(\epsilon^T C{\gamma^i}^\dagger\l-
			\epsilon^\dagger C\gamma^i\l^*\right)  \0\\
&&\delta A_{\mu}=-\left(\epsilon^\dagger\gamma^\mu\l-
			\l^\dagger\gamma^\mu\epsilon\right) \0\\
&&\delta \l=-\frac{1}{g^2}F_{\mu\nu}\gamma^{\mu\nu}\epsilon
	    -i\left[X_i,X_j\right]\gamma^{ij}\epsilon
	    -\frac{2i}{g}D_\mu X_i \gamma^\mu C \gamma^i \epsilon^*    
	    \label{esusy}
\ee
where, according to \cite{VN}, we consider the variables $\l^*$ and 
$\epsilon^*$ as independent from $\l$ and $\epsilon$, respectively. 
The superscript $^T$ represents the transpose matrix and $^\dagger$ stands
for $^{*T}$.

We look for solutions that preserve $\frac 14$ supersymmetry, by setting 
all fermions and all $X^i$, with $i=3,..,6$, to zero, and defining
$X=X^1+iX^2$ and $\bar X= X^\dagger$. The equations that define such solutions
are
\be
&&F_{v\bar v} + F_{w\bar w} -ig^2[X, \bar X]=0\label{ic1}\\
&&F_{v w}=0,\quad F_{\bar v \bar w}=0,\label{ic2}\\
&&D_{\bar v} X=0=D_{v}\bar X,
\quad D_w \bar X=0=D_{\bar w} X\label{ic3}
\ee
We will refer to the solutions of these equations as ic--instantons.
Analogous equations for ic--anti--instantons can be obtained by an 
anti--holomorphic involution. Similar equations were
previously discussed, in the context of ${\cal N}=4$ theory, 
for compact manifolds by \cite{VW}.

\subsection{Ic--instanton equations from self--duality in 8D}

Self--dual YM solutions in 4D are the well--known instantons.
Self--duality in 8D for the YM curvature is less known and was studied
a few years ago more as a curiosity than with the real aim at applying
it in physical problems, \cite{C,ward,FaNu,FuNi,BKS}. 
In components of the curvature the self--duality condition reads:
\be
&&F_{12}+F_{34}+F_{56}+F_{78}=0\0\\
&&F_{13}+F_{42}+F_{57}+F_{86}=0\0\\
&&F_{14}+F_{23}+F_{76}+F_{85}=0\0\\
&&F_{15}+F_{62}+F_{73}+F_{48}=0\0\\
&&F_{16}+F_{25}+F_{38}+F_{47}=0\0\\
&&F_{17}+F_{82}+F_{35}+F_{64}=0\0\\
&&F_{18}+F_{27}+F_{63}+F_{54}=0\label{selfduality}
\ee
The anti--self--duality equations are obtained from these by changing the sign
of the first entry of each one.

Let us reduce this system to 4D by keeping the dependence on $x^1,\ldots, x^4$
and dropping the dependence on the remaining coordinates. Let us introduce
the complex coordinates $v=\frac 12 (x^1+ix^2)$, $w= \frac 12 (x^3+ix^4)$, set
$A_7=A_8=0$ and
call $X=A_5-iA_6$. Then the system
(\ref{selfduality}) becomes:
\be
&&F_{v\bar v} + F_{w\bar w} +i[X, \bar X]=0\0\\
&&F_{vw}=0,\quad F_{\bar v \bar w}=0, \quad D_{\bar v} X=0=D_{v}\bar X,
\quad D_{\bar w} X=0=D_{w} \bar X\label{4dsys}
\ee
After an obvious rescaling, this is the system of equations found above as 
BPS equations, (\ref{ic1},\ref{ic2},\ref{ic3}). The connection of
these sets of equations with integrability is under study\footnote{We acknowledge
useful discussions with C.Constantinidis and L.Ferreira on this point}.

In the following section we would like to discuss solutions of (\ref{4dsys})
for which $X\neq 0$, i.e. ic--instantons.

\subsection{Ic--instanton equations: other derivations and covariant form}

It is worth spending a few more words on the equations (\ref{ic1}--\ref{ic3}).
We want to show here other ways in which they can be derived. This gives us
in particular the opportunity to write them in covariant form.  
First we notice that they are dimensional reduction of a single equation in 
6D K\"ahler manifold. The latter can be cast in covariant 
form using the K\"ahler form $\omega$:
\beq
\ast F = \omega \wedge F.
\label{6d}
\eeq
To see this, just use complex coordinates and write down its components; they 
can be reexpressed as
\beq
\omega \cdot F = 0, \qquad F^{(2,0)}=F^{(0,2)}= 0.
\label{dec}
\eeq
Now it is very easy to see that the dimensional reduction of this is nothing
but equations (\ref{ic1}-\ref{ic3}). One can
compare this with the situation in MST \cite{bbn1}, where the instanton 
equations (Hitchin's equations) are dimensional reduction of the single 
self-duality equation in 4d, which decomposes just as in (\ref{dec}). 

One can take an even more general point of view and look for solutions 
which preserve some fraction of supersymmetry in ten--dimensional 
SYM. This yields more general instanton equations. Writing
it explicitly in components, in 10d complex coordinates $(z_1 \ldots z_5)$, 
they look
\beq
\sum_{i=1}^5 F_{z_i, \bar z_i} = 0; \qquad F_{z_i, z_j}=0 \ \ \forall i<j,
\label{gen}
\eeq
which can be, again, rewritten in covariant form as $\ast F = \omega^3\wedge
F$. As particular cases, taking some $A_{z_i}$ vanishing,
we find the 8d equation $\ast F = \omega^2 \wedge F$ and the 
above mentioned 6d one. \\
If one dimensional reduces these to the dimension of interest, in
this case 4, one has also cases with more than one active scalar:
\be
&&F_{v\bar v} + F_{w\bar w} -ig^2\sum_{a=1}^3 [X_a, X_{\bar a}]=0\nonumber\\
&&F_{v w}=0,\quad F_{\bar v \bar w}=0, 
\qquad [X_a, X_b]=0, \quad \forall 1\leq a < b \leq 3
\label{active}\\
&&D_{\bar v} X_a=0=D_{v}X_{\bar a},
\quad D_w X_{\bar a}=0=D_{w} X_{\bar a}.\nonumber
\ee
These, however, do not represent new solutions as far as 
the problem we study in this paper is concerned.
In fact, anticipating
the discussion of the subsequent section, $[X_a, X_b]=0$
implies that $X_a = Y S \hat X_a S^{-1}Y^{-1}$, i.e. all $X_a$ are
diagonalized by the same matrix $YS$. This entails in particular that they
all have the same monodromy. Now, as we will see later,
each of these matrices $\hat X_a$ defines a covering of the base space, and
lifts to a holomorphic section of the trivial line bundle over the covering.
It follows that the $X_a$'s have to be multiple of one another. 
Therefore we can make a complex linear transformation and go back to   
the situation with just one active complex scalar.

Apart from this, the above equations open the way to interesting 
considerations, which are however outside the mainstream of this paper.
For this reason we limit ourselves here to some concise remarks.
Let us stick in particular to the 6d case. 
If $F$ is the curvature of a connection on a vector bundle, the quantity
$\int F\cdot \omega \ \nu$ (where $\nu$ is the volume form) is a topological
invariant, the {\sl degree}, which can be thought of as the intersection
$[c_1].[\omega^2]$ in homology and naturally generalizes the degree in
two dimensions ($=c_1$); one may easily show that a holomorphic line
bundle admits holomorphic sections iff its degree vanishes.
The second equation in (\ref{dec}) just means that our connection defines a
holomorphic structure on the vector bundle which is {\it integrable}; the
first one means that the bundle has degree zero. This condition fits
into a more general framework. A connection is called Hermitian-Yang-Mills if,
for some $\mu$, $\omega \cdot F = \mu\, {\rm Id}$. A theorem \cite{uhlyau} 
gives necessary and sufficient conditions for solutions to these equations to
exist in terms of a condition of {\it stability}. This generalizes the results  
known in 4d for ASD equations \cite{donkro}, in 2d (Narasimhan-Seshadri) and
for Hitchin equations.

\subsection{General properties of ic--instantons}

The system of equations (\ref{ic1},\ref{ic2},\ref{ic3}) has various types of 
solutions. Notice that, if we set $X=0$, (\ref{4dsys}) becomes the usual 
self--duality condition in 4D. Therefore the set of solutions of (\ref{4dsys}) 
will include in particular all the ordinary instantons of YM in 4D, compatible
with the topology of the base manifold.  
In principle we could
consider solutions with $X\neq 0$ and nonvanishing instanton number. However
the vector bundle $E$ is such that $c_2(E)$ is trivial, 
therefore we only consider solutions with vanishing instanton number.

Let us now compute the action of a configuration that satisfies (\ref{4dsys}).
Starting from (\ref{eSYM}) we get
\be
S_{inst}= \frac {1}{2} \int_{\cal X} d^2vd^2w&{\rm Tr}&\left( 
D_vXD_{\bar v}\bar X + D_v \bar XD_{\bar v}X+ D_wXD_{\bar w}\bar X +
D_w\bar XD_{\bar w} X  +\frac {g^2}{2} [X,\bar X]^2 \right.\0\\
&&\left.- \frac{1}{2g^2} (F_{v\bar v}^2 + F_{w\bar w}^2 - 2 F_{vw}F_{\bar v 
\bar w} - 2 F_{v\bar w}F_{\bar v w})\right) \label{instaction}
\ee
This can be rewritten as
\be
S_{inst}&=&S_{bulk}+ S_{boundary}\0\\
S_{bulk}&=&\int d^2v d^2w \,{\rm Tr} \left(
D_v\bar X  D_{\bar v} X +D_w\bar X  D_{\bar w} X +
\frac{1}{g^2} F_{vw}F_{\bar v\bar w}\right.\0\\
&&\left.\quad 
-\frac{1}{4g^2}(F_{v\bar v}+ F_{w\bar w} -i g^2[X,\bX])^2\right)\0\\
 S_{boundary}&=& \int d^2v d^2w\,
 {\rm Tr}\left(D_v(XD_{\bar v}\bar X) + D_w(XD_{\bar w}\bar X)+ 
\frac{1}{4g^2} d K(A_v,A_w)\right),\label{instaction'}
\ee
where $K$ is the Chern--Simons term corresponding to $F\wedge F$. More 
explicitly
$$
\left(F_{v\bar w}F_{\bar v w}-F_{vw}F_{\bar v \bar w} + F_{v\bar v}F_{w\bar w}
\right)d^2v d^2w = d K(A_v,A_w)
$$
where $d$ is the exterior derivative in 4D. One sees immediately that for 
an ic--instanton $S_{bulk}=0$. It is well--known that the Chern--Simons term
in $S_{boundary}$ is equal to the instanton number. Therefore, since in this 
paper we only consider solutions with instanton number 0, the Chern--Simons term 
does not contribute. The other term in $S_{boundary}$ is usually divergent for 
the
ic--instantons solutions and apparently one cannot attach any geometric meaning
to it. On the other hand it is very easy to get rid of it by simply saying
that our starting action is (\ref{eSYM}), in which the first two
terms have been modified to $-\frac 12 \Big(X^i\{D_v,D_{\bar v}\} X^i 
+X^i\{D_w,D_{\bar w}\} X^i\Big)$.
With these provisos the action of the ic--instantons considered in this paper
vanishes. A similar conclusion holds for ic--anti--instantons.

\section{Ic--instantons}

Our purpose is to find solutions $(A,X)$ of (\ref{ic1},\ref{ic2},\ref{ic3}). 
For  definiteness let us consider a concrete case, say 
${\cal X}= {\mathbb R}\times {\mathbb T}^3$.
The construction is parallel to the one carried out in \cite{bbn1,bbnt1}.
In the following we stick to the complex structure of the punctured sphere 
${\mathbb P}^1$ times ${\mathbb T}^2$, with local coordinates $v$ and $w$.
At times it is convenient to use the coordinate $z=e^v$. 
We start from the (simple) ansatz
\be 
A_v=i\d_v Y^\dagger (Y^{-1})^\dagger,\quad\quad
A_w=i\d_w Y^\dagger (Y^{-1})^\dagger, \quad\quad X=Y^{-1}MY\label{sansatz}
\ee
where $Y$ is a generic element in the complex group $SL(N,{\mathbb C})$
and $M$ specifies a branched covering of the base manifold. A more general
ansatz will be considered later on. As a 
consequence of (\ref{sansatz}) the equations $D_{\bar v} X=0=D_{ v}\bar X$ 
are equivalent to 
\be
\d_{\bar v} M=0=\d_{\bar w}M\label{holo}
\ee
which means that the matrix $M$ is holomorphic in $v,w$. Eq.{\ref{holo}}
guarantees that eqs.(\ref{ic3}) are satisfied.

The ansatz (\ref{sansatz}) is given in terms of two matrices, $Y$ and $M$. 
$Y$ will be called the {\it group theoretical factor}, while $M$ defines 
a general {\it branched covering} of the base manifold, i.e. a two dimensional 
complex manifold. The factor $Y$ will be 
discussed below, while branched coverings will be discussed later on.
For the time being let us give some essential information. 
Let us consider the polynomial
\be
P_X(y)=\det (y - X)=y^N+\sum_{i=0}^{N-1}y^ia_i\,,\0
\ee
where $y$ is a complex indeterminate. The equation
\be 
P_X(y)=0\label{spec} 
\ee 
can also be written as the matrix equation
\be 
X^N+a_{N-1}X^{N-1}+\cdots +a_0=0\,.\label{spectr} 
\ee 
A diagonalizable matrix, which is solution of eq.~(\ref{spectr}),
can always be cast in the canonical form
\be
M=\left(\matrix{-a_{N-1}& -a_{N-2}& \ldots & \ldots  & -a_0\cr
		  1      & 0       & \ldots & \ldots  & 0  \cr
		  0       & 1       &  0  & \ldots  & 0 \cr
		  \ldots	   & \ldots     &  \ldots& \ldots  & 
		  \ldots \cr		
                  0	    & 0       & \ldots & 1    & 0 \cr}\right).\label{M}
\ee

Due to~(\ref{holo}), we have $\d_{\bar v}a_i=0=\d_{\bar w}a_i$, which means 
that the set of functions $\{a_i\}$ are holomorphic in $v,w$,
although they are
allowed to have poles at $z=0$ and $z=\infty$. The point is that,
as we shall see in many examples, Eq.(\ref{spec})
identifies in the $(y,z,w)$ space a complex 2--manifold (a surface) $\Sigma$, 
which is an 
N--sheeted branched covering of the base manifold. The explicit form of 
the covering is given by the set
$\{x^{(1)}(z,w),\dots,x^{(N)}(z,w)\}$ of eigenvalues of $X$. Each
eigenvalue spans a sheet. The projection map to the base cylinder
${\cal X}$ will be denoted $\pi:\Sigma \to {\cal X}$. The divisor
(complex 1--submanifold) where
two eigenvalues coincide is the branch locus. We can also define branch cuts:
they are 3d manifolds that connect disconnected components of the branch locus.

We stress that the covering is independent of the coupling $g$.

\subsection{Explicit construction of ic--instantons}

The aim of the present subsection is to construct the group theoretical factor
corresponding to the most general covering. The construction is close to the
one in \cite{bbnt1}, so we will be brief.  
 
Let us recall our ansatz (\ref{sansatz}).
The group theoretical factor $Y$ takes values in the complex group
$SL(N,{\mathbb C})$, while the matrix $M$ determines the branched
covering. The dependence on the Yang-Mills coupling
constant $g$ is contained in the $Y$ factor, while $M$ does not depend
on $g$. We set $Y= KL$
where $L$, {\it the dressing factor}, is expected to tend to 1 in the strong
coupling limit outside the branch locus, while $K$ is a
special matrix, independent of $g$, endowed with the property that
$K^{-1} M K$ and $K^\dagger M^\dagger (K^\dagger)^{-1}$ are
simultaneously diagonalizable. 

It is well-known,
\cite{wynter}, that the matrix $M$ can be diagonalized
\begin{eqnarray}
M = S \hat M S^{-1}, \quad\quad \hat M = {\rm Diag}(\lambda_1,\ldots, \lambda_N)
\end{eqnarray}
by means of the following matrix $S\in SL(N,{\mathbb C})$:
\be
S= \Delta^{-{1\over N}}\left(
	\matrix{\l_1^{N-1}&\l_2^{N-1}& \ldots & \ldots &\l_N^{N-1}\cr
		\l_1^{N-2}&\l_2^{N-2}& \ldots & \ldots  &\l_N^{N-2}\cr
		\ldots	& \ldots  & \ldots & \ldots  &  \ldots \cr
                1	& 1	  & \ldots & \ldots  & 1 \cr}\right),\label{S}
\ee
where 
\be
\Delta = \prod_{1\leq i<j\leq N}(\l_i - \l_j)\,.\label{Delta}
\ee
$\Delta$ vanishes whenever two 
eigenvalues coincide. Two coincident eigenvalues define a component of
the branch locus of
the covering. Going around a branch locus and crossing a branch cut in the
$v,w$--plane, produces a reshuffling of the eigenvalues that can be
represented via a monodromy matrix $\Lambda$: $\hat M\to \Lambda \hat M 
\Lambda^{-1}$. Correspondingly we have $ S\to S\Lambda^{-1}$, so that
the single--valuedness of $M$ is preserved. 

The explicit construction of $K$ and $L$ in the general case is given in 
\cite{bbnt1} and will not be reported here. The qualitative features 
are as follows.
First one introduces a monodromy--invariant $K$ such that $K^{-1}S=U$ be 
unitary. To this end one sets $K= \sqrt{SS^\dagger}$ and easily verifies
that $U$ is unitary. As it turns out, $K$ may have
singularities at the points of ${\cal X}$ where any two eigenvalues of
$M$ coincide, i.e. at the branch locus of the spectral
covering (the elements of $K$ contains as factors fractional powers 
of $|\Delta|$). Therefore $K^{-1}MK$ is in general singular at these
points. That is why we must introduce into the game a new monodromy
invariant matrix $L$, with the purpose of canceling the singularities of 
$K^{-1}MK$ in such a way that $L^{-1}K^{-1}MKL$ be smooth and satisfy
(\ref{ic1},\ref{ic2}). Let us denote again by $\phi$ the generic entry of $L$. 
For (\ref{ic1}) to be satisfied $\phi$ must 
satisfy, \cite{bbnt1}, an equation of the WZNW type with the following
general structure
\be
(\d_v\d_{\bar v}+\d_w\d_{\bar w})\phi + ... \sim(\d_v\d_{\bar v}+ 
\d_w\d_{\bar w}) \ln |\Delta|= {\pi}\Big(
\d_v \Delta \d_{\bar v }{\bar\Delta}+ \d_w \Delta \d_{\bar w }{\bar\Delta}\Big)
\delta({\Delta})\,,
\label{dress1}
\ee
where dots represent all the other terms, which are irrelevant in the 
cancellation of singularities. In some equations (but not in all) the 
coefficients in front of the delta--function terms may vanish. This term
has support at the zeroes of $\Delta$, i.e. at the branch locus.
The equation $F_{vw}=0$ in (\ref{ic2}), on the other hand, does not give rise 
to delta function terms:
\be
\d_v\d_w\phi+...=0\label{dress2}
\ee 
Let us refer to the above equations collectively as 
the `dressing equations'.

By construction $K$
is independent of $g$ while $L$ does depend on $g$. One can show that
in fact $L\to 1$ as $g\to \infty$, outside the zeroes of the discriminant. 
Let us present a simple argument in this sense.

The solution $X$ exists with the required properties only if
the `dressing equations' admit solutions that vanish at $v=\pm
\infty$ (to this end, of course, we have to exclude possible branch locus
components at $t=\pm \infty$ from the right hand side of eq.(\ref{dress1}). 
To our best knowledge, not much is known in the literature
concerning the existence of such solutions. Based on the analysis
of \cite{bbn2}, we assume that the
`dressing equations' do admit solutions that vanish at $v=\pm
\infty$. Once one assumes this, it is rather easy to argue, on a
completely general ground, that in the strong coupling limit, $g \to
\infty$, such solutions vanish outside the zeroes of the discriminant.
The argument goes as follows. Consider a candidate solution of
(\ref{ic1},\ref{ic2}) in which $\phi=0$ outside the zeroes of the
discriminant, for all the $\phi$'s. Then, there, $L=1$, and $X= K^{-1}MK$.  
As noted previously, in such a situation $[X, \bar X]=0$,
since both $X$ and $\bar X$ are simultaneously diagonalized by the
matrix $U=K^{-1}S$. Now we have to show that also $F_{v\bar v}$ and
$F_{w\bar w}$ vanish 
outside the zeroes of the discriminant if $L=1$. In fact when $L=1$, 
\be
A_{\bar v}= -i K^{-1}\d_{\bar v} K=-i (K^{-1}SS^{-1})\d_{\bar v} (SS^{-1}K)
= -i U (\d_{\bar v} +
\tilde A_{\bar v})U^{-1}\,,\0
\ee
where $ \tilde A_{\bar v} = S^{-1}\d_{\bar v} S$. But $\d_{\bar v} S\equiv 0$ 
due to holomorphicity of the eigenvalues of $M$. Therefore $F_{v\bar v}=0$.
The same can be done for $A_w$, therefore $F_{w\bar w}=F_{vw}=0$.
In conclusion
(\ref{ic1},\ref{ic2}) is identically satisfied by the ansatz $L=1$ outside
the zeroes of the discriminant. Since the solutions are uniquely
determined by their boundary conditions, we can conclude that, as
$g\to \infty$, the only solution of the dressing equations outside the
zeroes of the discriminant, is the identically vanishing solution. We
infer from this argument that the solutions of the dressing equations
for large $g$ are concentrated around the branch locus and become
more and more spiky as $g$ grows larger and larger. Therefore the
matrix $L$ has the properties we expect.

The previous argument hinges on the occurrence that, as $g=\infty$, we have 
both $[X^\infty, \bar X^\infty]=0$ and 
$F_{v\bar v}^\infty=F^\infty_{w\bar w}=0$ (the superscript $^\infty$ obviously 
represents the strong coupling value of a field). Other types of solutions 
can be envisaged, see below and \cite{bbnt1}. 

\subsection{Generalized ic--instantons}

In this paper we will have to take into consideration more general solutions 
than those just studied. Instead of (\ref{sansatz}) let us start from
\be 
A_v=i{\cal D}_v Y^\dagger (Y^{-1})^\dagger,\quad\quad
A_w=i{\cal D}_w Y^\dagger (Y^{-1})^\dagger, \quad\quad X=Y^{-1}MY\label{gansatz}
\ee
where $Y$ is as before and the covariant derivative ${\cal D}$ is relative
to a connection ${\cal A}$ which commutes with $M$. As a 
consequence of this, we have again that the equations 
$D_{\bar v} X=0=D_{ v}\bar X$ imply (\ref{holo}). Moreover, the connection 
${\cal A}$ is diagonalized by $S$ and 
\be
{\cal A}= S \hat A S^{-1}\0
\ee
Since any solution $A, X$ must be smooth, the monodromy
of $\hat A$ must be the same as the monodromy of $\hat M$, i.e. going around a 
branch locus produces a reshuffling of the eigenvalues that can be
represented via the same monodromy matrix $\Lambda$: $\hat A\to \Lambda \hat A 
\Lambda^{-1}$.

The construction of such ic--instantons carries through as before. The only 
remarkable difference is that in the strong coupling limit the connection 
${\cal A}^{\infty}= U\hat A U^{-1}$ does not evaporate into a pure gauge as before. Not only 
do we have a covering described
by $\hat X$, but also a connection $\hat A$ valued in the Cartan subalgebra. 

In the strong coupling limit, instead of (\ref{ic1}), in this case we find
\be
&&[X^\infty,\bar X^\infty] =0\0\\
&&F_{v\bar v}^\infty+F_{w\bar w}^\infty =0\label{ginst}
\ee
i.e. we obtain a non--trivial self--dual connection. Of course we can do the 
same with anti--self--dual ic--instantons and obtain strong coupling
antiself--dual connections

An important proviso: a basic condition for us to call all the above
solutions {\it ic--instantons} is that $[X,\bar X]\neq 0$ for finite $g$: 
only in this case do they represent interpolating solutions
between genuine initial and final brane configurations (see below).

\section{Spectral coverings and scattering}

In the previous section we have seen that in the strong coupling limit
any ic--instanton reduces to a branched covering of the base manifold ${\cal X}$.
In this section we analize a few general facts and examples of branched
coverings of 4--manifolds (without any illusion of completeness). 
We will see that, in parallel to what happens in MST, such coverings 
may describe scatterings of D3--branes. The idea is simple. 
The ic-instantons present in our theory describe four--manifolds of various 
topologies, which cover various base spaces; the latter are topologically of the 
form $\rr \times M_3$, and so we may define slices of the covering at 
constant time. In general our ic--instantons at $t=-\infty$ are represented 
by a disjoint union of 3-manifolds of various topology and at $t=+\infty$ by 
another (in general different) disjoint union. Now, we interpret the $t=-\infty$
configuration as a set of incoming 3--branes and the $t=\infty$ one as a set of
outgoing 3--branes. Any ic--instanton interpolates between two such asymptotic
configurations. If we want to describe a given scattering process we 
will choose, among all the ic--instantons, those with the
given asymptotic structure, i.e. whose slices at $t \to \pm \infty$ correspond
to the assigned unions of 3--manifolds.

We are therefore faced with two classification problems: 1) given
a base manifold ${\cal X}$, classifying all possible branched coverings;
2) analyzing the effect of a change of base manifold.

From a path--integral point of view it is clear from the example of MST that
we have to sum over all the branched coverings of 1), which means a discrete 
sum over topologies and an integration over moduli. We will argue later
on that we have perhaps to sum also over different base manifolds.

In this paper we will actually limit ourselves to analyzing two special cases 
of scattering topology, with the purpose of illustrating these two problems. 
Let us, however, point out 
first a general result. We will denote by $\Sigma$ the complex surface
associated to a given instanton and by $\pi$ the projection
$\pi:\Sigma\to {\cal X}$. As we will see later on any instanton relevant for a 
given process will contribute to the path integral a term proportional to 
$g^{-\chi}$, where $\chi$ is the Euler characteristic of the 
4-manifold ${\Sigma}$. To compute $\chi$, we can triangulate the
manifold in such a way that it gives a triangulation of the branch locus as
well. Doing so we obtain the result
\beq
\chi_{\Sigma} = N \chi_{\cal X} - \sum (r_i -1) \chi_{R_i},
\label{euler}
\eeq
where $r_i$ are the ramification orders at the branch loci $R_i$, $N$ is 
the order of the covering and ${\cal X}$ is the base. We recall that in our case 
$\chi_{\cal X}=0$.

\subsection{Scattering of $S^3$--branes}

The natural choice of base space is in this case $\rr\times S^3$.
 In this case, the complex structure we can take is obviously 
${\mathbb C}^2 - \{(0, 0)\}$. In terms of the complex coordinates $(v,w)$
introduced above the time is given by $e^t =\sqrt{|v|^2 +|w|^2}$.
The characteristic polynomial depends on both coordinates; the
branch locus is a curve in ${\mathbb C}^2 - \{(0, 0)\}$, given by some equation
$\Delta(v,w)=0$, and its slices at constant time are generically unions of 
$S^1$. 

Let us consider first the asymptotics at $t=-\infty$. In a generic situation
we expect the inverse image of $S^3$ under $\pi$ to be a disjoint union of
$N$ copies of $S^3$. But of course we are interested in less trivial 
asymptotic configurations. This is so if the point at $t=-\infty$ belongs
to the branch locus. In such a situation we can use, for instance, 
the well-understood theory
that relates germs of plane curves (i.e. local forms of equations in $\cc^2$)
 to knots and links embedded in a small $S^3$ around the origin \cite{brie}.
Let us review some of those results. First of all, it is obvious that if the
equation $\Delta =0$ has a constant term, a small sphere does not intersect the branch
locus. Apart from this trivial case, the rule is that each component of
$\Delta$ as a polynomial corresponds to an $S^1$, and the multiplicity of
intersection of two components is exactly the linking number of the
corresponding $S^1$. As for each single component, its local Puiseux expansion
encodes the knot type of the corresponding $S^1$'s.

Once we have understood the structure of the branch locus $R$, given 
that the number of sheets of the covering is $N$, we need to know
the action of the first homotopy group $\pi_1(S^3 - R)$ on the discrete fiber
(the monodromy). From these data we can reconstruct
topologically the covering space. Suppose, for definiteness, that the covering 
is totally branched along a knot, and the monodromy is the generator of the
cyclic group $\zz_N$ ({\it cyclic covering}). 
An easy case we may analyze is that in which the knot is
trivial; think indeed the base $S^3$ as
$\rr^3 \cup \infty$, choose a line in it as branch (from the point of view of
$S^3$ it is a circle) and construct the branched covering
as usual, attaching in sequence $N$ copies of $\rr^3$ along the $S^1$ (this is
the same as the usual picture of a 2d branched covering, translated along one
more spatial direction). The covering space is again an $\rr^3$ with a point at
infinity, so it is an $S^3$ as well. 

To describe more complicated cases, there is a beautiful theory relating
branched coverings, knots and surgeries \cite{prasos}. Surgery is a
technique that allows one to obtain any 3--manifold from a sphere, cutting a
solid 2--torus constructed along a knot, and regluing it in a different
way. By applying it to the simple branched covering $S^3 \to S^3$ we just
described, one may induce other branched coverings $M_3 \to S^3$, 
branched along knots. In fact, one may show that {\sl any} $M_3$ can be
obtained as a covering of $S^3$, with branch along a knot. The problem is,
however, that the knots obtainable as branches from complex geometry are not
of general type; they are called {\it iterated torus knots}. Therefore
we see that the base $\mathbb R \times S^3$ may give rise to many different 
topologies
at $t=-\infty$, but we have no guarantee that it gives rise to any desired 
topology for the incoming branes.

In some cases, there is a further method to understand the topological structure
of the scattering surface. If the equation of the covering $p(y, v, w)=0$
is homogeneous in 
$(y, v, w)$, and if the coefficient of $y^N$ is 1, we may map homeomorphically 
the solutions $\{ |v|^2 + |w|^2 = c, p=0 \}$ above the 3--sphere to solutions
$\{ |y|^2 + |v|^2 + |w|^2 = c, p=0 \}$, exploiting the fact that 
$ p(y, v, w)= 0 \Leftrightarrow p(\lambda y, \lambda v, \lambda w)=0$. Think of
this as the stereographic projection from a cylinder to the sphere inscribed
in it. In this case it is simpler to understand the topology: 
it is the intersection of a homogeneous equation in $\cc^3$ with the sphere $S^5$.
The homogeneous
equation can be read as an equation  in $\pp^2$; $S^5$ can be thought of as
the $U(1)$ bundle inside ${\cal O}_{\pp^2} (-1)$, and so we get in this case
that our brane has the topology of a $S^1$ bundle over a Riemann surface (it
is, in fact, just the $S^1$ inside the line bundle which embeds the Riemann
surface in $\pp^2$).\\
For quasi--homogeneous equations, a similar projection can be done; an
analysis in terms of weighted projective spaces is however less straightforward,
and one has to resort to other methods, \cite{kauf}.

The analysis carried out so far only concerns asymptotic branes
(the $t=+\infty$ case is analogous to the $t=-\infty$ one). At finite time it
is still true that, given the structure of the branch locus, we can 
single out the intermediate configurations but the analysis is in general
more difficult. In general, what happens is that the initial 
branes will join and split in branes of the same or different topologies. 
As an example, consider the polynomial $y^N = v^2- w^2 - e^{t_0}$.
The branch locus is absent for $t<t_0$, and an unknotted $S^1$ for $t>t_0$.
So, by the above discussion, this describes $N$ spheres which join to form
one.

Up to now, we tried to describe a scattering of spheres by considering the
most natural base $\rr \times S^3$, and we found scattering states of very
general topology. But to be complete, we should describe other bases, and 
see whether
there are instantons with spheres as asymptotic structure. As we said, this
means that there should be a covering $S^3 \to M_3$ coming from the
restriction at fixed time of a complex branched covering. Just as the
base $\rr \times S^3$ contributes to scattering of all manifolds, other bases
may contribute to the scattering of $S^3$s. One would have to extend the
theory we cited above \cite{prasos} to base 3--manifolds different from
$S^3$. We will not try this here.

\subsection{Scattering of ${\mathbb T}^3$--branes.}

Also in this case we begin with a base space $\rr \times {\mathbb T}^3$. There
are indeed complex structures on it: think of this base as $\rr^4 / \Lambda$,
where $\Lambda$ is a 3--lattice spanned, say, by $v_1, v_2, v_3$. Now take the
standard complex structures on $\rr^4 = \cc^2$: we obtain, varying $v_i$,
different complex structures. To be more precise, they are really different
only modulo the action of $GL(2, \cc)$ and modular transformations. This
general setting, however, yields results not different from those obtained 
by taking the simplest choice among them: thinking of the base space as 
$ (\pp^1 - \{ 0 , \infty\}) $ times an elliptic curve 
${\cal C}$. 
We call $z$ the coordinate on
the first factor (time is given by $e^t = |z|$) and $w$ the one on
${\cal C}$.

The coverings are defined by the characteristic polynomial $P_X$, whose
coefficients $a_i$ are holomorphic in $z$ and $w$ by \ref{holo}. 
As functions of 
$z$, they are just meromorphic functions on $\pp^1$, with poles in the
excluded points $0$ and $\infty$; as functions of $w$, they 
are constant. The resulting coverings are
very simple: for each fixed $z$, the covering space is nothing but a disjoint
union of 2-tori (the eigenvalues are constant in $w$); so the process is of
the type (scattering of strings)$\times T^2$, where the first factor is
exactly what was already examined in MST \cite{bbn1,bbn2,bbnt1}. This simply
means that 3-tori are really scattering just along one of their dimensions.
Since in this case the branches $B_i$ are tori, by (\ref{euler}) these processes 
all have $\chi =0$, and therefore contribute only to the zero-th order term in 
$1/g$ in the path integral.

This is the simplest possibility; but again we have to consider contribution
from other bases, with branched or unbranched coverings ${\mathbb T}^3 \to M_3$. 
The first idea is to use $\rr\times S^3$, which
 we have considered above. We are not sure that there is actually a covering 
yielding ${\mathbb T}^3$; the technique to construct it would be to analyze 
coverings along
iterated torus knots, coming from equations $\Delta (v, w)=0$, and then
construct a holomorphic covering having $\Delta$ a discriminant. Similar
analysis should be done for other bases.

\section{Expansion about a classical solution}
 
Our purpose in this section is to expand the action about a classical
ic--instanton solution. For definiteness
we choose an instanton rather than an anti--instanton, but everything can be 
repeated for the latter. The analysis is along the lines of \cite{bbn2}, but
there are important differences which we will try to emphasize while going
rapidly through the repetitive aspects.

As a first step let us analyze the background part. 
The dependence on the coupling is entirely contained in the 
factor $L$. We have seen that in the strong coupling limit 
$L\to 1$ outside the branch locus of the covering. 
Since here we are interested in expanding 
the action (\ref{eSYM}) in inverse powers of $1/g$, and actually in singling 
out the dominant term in this expansion (see below), we will consider the 
action (\ref{eSYM}) around a given classical solution stripped of the above 
dressing factor, and exclude from the integration region the branch locus on 
the base manifold, ${\cal X}$. In other words we will consider from now on 
the action (\ref{eSYM}) in which the relevant $Y$ is replaced by 
$K$ and the integral extends over ${\cal X}_0$ which is the 
initial ${\cal X}$ from which small tubular neighborhoods have 
been cut out around the branch locus. Said otherwise, we introduce in our 
integrated action a regulator (which will eventually be removed).

After getting rid of the dressing factor, the classical background 
configuration is specified by $X^\infty$ and 
$A^\infty$ (see section 3.2). 
As expected, this configuration is singular exactly at the branch locus. 
We have seen that $M= S\hat M S^{-1}$.
$\hat M$ is the matrix of eigenvalues of $M$ and of $X$, so we denote 
it equivalently by $\hat X$. In the strong coupling limit 
$X \to U \hat X U^{-1}$, where  $U=K^{-1} S$  is a unitary matrix
and therefore simultaneously diagonalizes $X$ and $\bar X$.  
Corresponding to $\hat X$ we have $\hat A_v, \hat A_w$.

$U$ is finite in ${\cal X}_0$.
Therefore, with a gauge transformation, we can remove it
from the action defined in ${\cal X}_0$. This leads us to 
\begin{itemize}
\item $\hat X$ and $\hat A$ diagonal,
\end{itemize} 
for the classical background in the strong coupling limit. 
 
Let us return now to the bosonic action (\ref{eSYM}) (the fermionic part will 
be analyzed later on) with the above understanding of the background part.
To extract the strong coupling effective theory, we first rewrite 
the action in the following useful form
\be
S^{(b)}&=&\int d^2v d^2w \,{\rm Tr} \left(
D_vX^I D_{\bar v} X^I +D_wX^I D_{\bar w} X^I
-\frac{g^2}{2}[X^I,X^J]^2
-g^2[X^I,X][X^I,\bX]\right.\0\\
&&\left.+ D_v\bar X  D_{\bar v} X +D_w\bar X  D_{\bar w} X +
\frac{1}{g^2} F_{vw}F_{\bar v\bar w}
-\frac{1}{4g^2}(F_{v\bar v}+ F_{w\bar w} -i g^2[X,\bX])^2\right)
,\0
\ee
where $I= 3,...,6$. 
We now expand the action around a generic ic--configuration as follows 
\be
\Phi=\Phi^{(b)}+\phi^{\mathfrak t}+\phi^{\mathfrak n}\equiv 
\Phi^{(b)}+\phi\equiv \Phi^\circ +\phi^{\mathfrak n}\,,
\ee
where $\Phi^{(b)}$ is the background value of the field at infinite
coupling, $\phi^{\mathfrak t}$ are the fluctuations along the Cartan directions
and $\phi^{\mathfrak n}$ are the fluctuations 
along the complementary directions in the Lie algebra ${\mathfrak u}(N)$. 
In the following we suppose we have carried out
the operation described above and by background value we refer to
the diagonal representation.

The expansion of the action starts with quadratic terms in the fluctuations
and ${\hat A}$ drops out from all the terms, except from the kinetic 
energy term of the (diagonal) Yang--Mills field. 
To simplify the subsequent formulas we will drop $\hat A$
for the time being and resume it later on.

To proceed further let us fix the gauge.
We use, in the strong coupling limit, the following gauge--fixing term
\be
S_{gf}=\frac{1}{4\pi g^2}\int d^2vd^2w~{\rm Tr}~{\cal G}^2
\ee
where
\be
{\cal G}=D^\circ_v a_{\bar v} +D^\circ_{\bar v} a_v+
 D^\circ_w a_{\bar w} +D^\circ_{\bar w} a_w + i g^2 ([X^\circ, \bar x]
+ [{\bar X}^\circ, x])+ 2i g^2 [ X^{\circ I}, x^I]  \label{gf}\,,
\ee
and $D^\circ$ is the covariant derivative with respect to $A^\circ$.
Next we introduce the Faddeev--Popov ghost and antighost fields $c$ and 
$\bar c$ and expand them like all the other fields and 
add to the action the corresponding Faddeev--Popov ghost term
\be
S_{ghost}= -\frac{1}{2\pi g^2} \int d^2v d^2w ~{\rm Tr}~\left(\bar c 
\frac {\delta {\cal G}}{\delta c}c \right)\,,
\ee
where $\delta$ represents the gauge transformation with parameter $c$.

At this point, to single out the strong coupling limit of the action,
we rescale the fields in appropriate manner. Precisely, we redefine our fields 
as follows
\be
A_v = g a_v^{\mathfrak t} + a_v^{\mathfrak n} ,\quad
A_w = g a_w^{\mathfrak t} + a_w^{\mathfrak n}, \quad
X= \hat X + x^{\mathfrak t} + \frac {1}{g} x^{\mathfrak n},\quad
X^I = x^{I{\mathfrak t}} + \frac{1}{g} x^{I{\mathfrak n}} \0
\ee
and likewise for the conjugate variables. For the ghosts we set
\be
c= g c^{\mathfrak t} + \sqrt g  c^{\mathfrak n}, \quad
\bar c= g \bar c^{\mathfrak t} + \frac {1}{\sqrt g} \bar c^{\mathfrak n}\,.
\0
\ee
 
After these rescalings the action becomes
\be
S^{(b)}=S^{(b)}_{sc}+S^{(b)}_{\mathfrak n}+o\left(\frac{1}{\sqrt{g}}\right)\0\,,
\ee
where
\be
S^{(b)}_{sc}&=&
\int_{{\cal X}_0} d^2vd^2w \,~{\rm Tr}~ \left[
\d_v x^{I{\mathfrak t}} \d_{\bar v} x^{I{\mathfrak t}}
+\d_w x^{I{\mathfrak t}} \d_{\bar w} x^{I{\mathfrak t}} 
+\d_vx^{{\mathfrak t}} \d_{\bar v} \bar x^{{\mathfrak t}} 
 +\d_w x^{{\mathfrak t}} \d_{\bar w} \bar x^{{\mathfrak t}} 
 \right.\0\\
&&\left.+\d_v \bar c^{{\mathfrak t}} \d_{\bar v} c^{{\mathfrak t}} 
+\d_w \bar c^{{\mathfrak t}} \d_{\bar w} c^{{\mathfrak t}}
+\d_v a^{\mathfrak t}_{\bar v}\d_{\bar v} a^{\mathfrak t}_v 
+\d_w a^{\mathfrak t}_{\bar w}\d_{\bar w} a^{\mathfrak t}_w 
 +\d_v a^{\mathfrak t}_{\bar w}\d_{\bar v} a^{\mathfrak t}_w 
+\d_w a^{\mathfrak t}_{\bar v}\d_{\bar w} a^{\mathfrak t}_v 
\right]\label{scaction}
\ee
$S^{(b)}_{\mathfrak n}$ is the purely quadratic term in the 
$\phi^{\mathfrak n}$ fluctuations. Let us see this in detail.

$S^{(b)}_{\mathfrak n}$ has the form
\be
S_{\mathfrak n}=  \int d^2vd^2w ~ {\rm Tr}~\left[ \bar 
x^{\mathfrak n}{\cal Q} x^{\mathfrak n}+  
x^{I{\mathfrak n}}{\cal Q} x^{I{\mathfrak n}}+
 a_{\bar v}^{\mathfrak n}{\cal Q} a_v^{\mathfrak n}+
a_{\bar w}^{\mathfrak n}{\cal Q} a_w^{\mathfrak n}+
\bar c^{\mathfrak n}{\cal Q} c^{\mathfrak n}
\right]\label{Qn}\,,
\ee
where 
\be
{\cal Q} = {\rm ad}_{{\bar X}^\circ}\cdot {\rm ad}_{ X^\circ}+
{\rm ad}_{a_{\bar v}^{\mathfrak t}}\cdot {\rm ad}_{a_v^{\mathfrak t}} +
 {\rm ad}_{a_{\bar w}^{\mathfrak t}}\cdot {\rm ad}_{a_w^{\mathfrak t}} +
 {\rm ad}_{x^{I{\mathfrak t}}}\cdot {\rm ad}_{x^{I{\mathfrak t}}}\0
 \ee

There are no zero modes involved; therefore the integration gives a certain
power of the determinant of ${\cal Q}$. This has to be compared with the 
fermionic part of the action. So let us look at the latter. After the rescaling 
$\l=\l^{\mathfrak t}+\frac{1}{\sqrt g}\l^{\mathfrak n}$, we have analogously
\be
S^{(f)}=S^{(f)}_{sc}+S^{(f)}_{\mathfrak n}+o\left(\frac{1}{\sqrt{g}}\right)\0\
\ee
where, ,
\be
S^{(f)}_{sc}&=&\int_{{\cal X}_0} d^2vd^2w\left[
-2\left( {\l_1^{\mathfrak t}}^* \d_{\bar v}{\l_1^{\mathfrak t}}
+{\l_2^{\mathfrak t}}^* \d_v{\l_2^{\mathfrak t}}\right)
-2\left( {\l_1^{\mathfrak t}}^* \d_{\bar w}{\l_2^{\mathfrak t}}
-{\l_2^{\mathfrak t}}^* \d_w{\l_1^{\mathfrak t}}\right)\label{scfermaction}
\right]
\ee
The fermionic off--diagonal fluctuations contribute quadratically in the 
following way. We  
arrange the $\l_\alpha^{\mathfrak n}$ and $\l_\alpha^{{\mathfrak n}*}$ 
in a unique ``spinor'' 
${\psi^{\mathfrak n}}^T=({\l_1^{\mathfrak n}},{\l_2^{\mathfrak n}},
{\l_1^{\mathfrak n}}^*,{\l_2^{\mathfrak n}}^*)$,
\be
S^{(f)}_{\mathfrak n} =
\int  d^2vd^2w~{\psi^{\mathfrak n}}^T {\cal A}~\psi^{\mathfrak n}~,
\ee
where
\be
&&{\cal A}=\left(\matrix
{0  & {\gamma^i}^\dagger{\rm ad}_{X_i^{\circ}} & 
-i{\rm ad}_{a_{\bar v}^{\mathfrak t}}& -i{\rm ad}_{a_w^{\mathfrak t}}\cr
 -{\gamma^i}^\dagger{\rm ad}_{X_i^{\circ}}&0&
i{\rm ad}_{a_{\bar w}^{\mathfrak t}}&-i{\rm ad}_{a_v^{\mathfrak t}}\cr
-i{\rm ad}_{a_{\bar v}^{\mathfrak t}}&i{\rm ad}_{a_{\bar w}^{\mathfrak t}}&0& 
-\gamma^i{\rm ad}_{X_i^{\circ}}\cr
    -i{\rm ad}_{a_w^{\mathfrak t}}& -i{\rm ad}_{a_v^{\mathfrak t}}& 
\gamma^i{\rm ad}_{X_i^{\circ}}&0  \cr}\right)
\ee

Now let us observe that the components of this matrix commute with respect to
 the action of the adjoint, and to the $SU(4)$ indices, so that we can 
directly compute the determinant looking at it as a $4\times 4$ matrix. 
Taking into account the $SU(4)$ and Lorentz indices, we get
\be
{\rm Det}{\cal A} 
=\left({\rm Det} {\cal Q}\right)^8
\ee
 
As this is precisely the determinant provided by the path integration on 
fermions, we now have to compare it with the bosonic one. This last turns out 
to be 
$\left({\rm Det}{\cal Q}\right)^{-8}$, obtained counting 6 scalars plus 4 gauge 
bosons 
minus 2 ghosts, and taking into account that the number of bosons too has been 
doubled, as an effect of the Wick rotation. So the final net contribution of 
the ${\mathfrak n}$ fields to the partition function is 1.

As it was pointed out in \cite{bbn2}, each separate entry of the diagonal 
matrix fields appearing in (\ref{scaction}) is not a true free field, as it is 
not single--valued. However each diagonal matrix field defines a unique 
(single--valued) field on the covering surface $\Sigma$ of ${\cal X}$ 
(see Appendix).
For example the matrices $x^{I{\mathfrak t}}$ represent scalar fields
${\bf x^I}$, the matrix $a^{\mathfrak t}$ represents a one--form field
${\bf a}$ on $\Sigma$ and so on. A boldface letter will be henceforth the
hallmark of a well--defined bosonic field on $\Sigma$. As for $\lambda$
its global existence on $\Sigma$ understands that the latter is a spin manifold. 

In conclusion the strong coupling theory represents
a free $U(1)$ gauge theory with matter on $\Sigma$:
\be
S= \frac{1}{2} \int_\Sigma d^4\xi \left( \frac 12 \d_\mu {\bf x}^i 
\d^\mu {\bf x}^i +\frac 12 \d_\mu{\bf a}_\nu  \d^\mu{\bf a}^\nu +
\frac 12   \d_\mu {\bf {\bar c}}\d^\mu {\bf c}- 
\frac 12 \lambda^\dagger \gamma^\mu
\d_\mu \lambda\right)\label{finalac}
\ee
where $\xi$ are local coordinates on $\Sigma$ (for example, $z$ and $w$).
The expression of the strong coupling (\ref{finalac}) is only symbolic. It is 
in fact strictly valid only if $\Sigma$ is a flat manifold, in which case
we recover full ${\cal N}=4$ supersymmetry. But of course
in general $\Sigma$ will not be flat. In the non--flat cases (\ref{finalac})
will only hold outside a neighborhood of the ramification locus, in which
the curvature is concentrated. The problem of course is not how to extend the 
action (\ref{finalac}) in such a way as to incorporate a non--trivial metric, 
which is straightforward, but rather how to do it in a supersymmetric way,
so as to obtain an ${\cal N}=4$ supersymmetric theory. This problem is analogous
to the covariant formulation of Green--Schwarz superstring theory on a generic
Riemann surface, met in MST. The difficulty of such problems stems from the fact 
that, at first sight, it would seem inevitable to introduce supergravity on the
world--volume in order to guarantee supersymmetry. However this is not necessary.
In fact both these problems, as as well as other similar
problems concerning D--brane actions embedded in space--time, have been solved
using the superembedding principle. An essential role is played by
$\kappa$ symmetry,
and the above mentioned difficulty is overcome by pulling back the (possibly 
trivial) metrics and gravitinos from the ambient space, which are therefore
non--dynamical. All this fits very 
well in our approach, and we limit ourselves to relying on the literature:
the action will be an extension of (\ref{finalac}) to include the branch locus
-- possibly substituting the SYM action with the corresponding DBI one. 
In our specific case we have in mind \cite{BST} (for a review of this and 
related problems, see \cite{sorokin}).  

We remark that (\ref{finalac}) contains the fields which are expected to live 
on a D3--brane and it is itself the low energy and low curvature action for 
a D3--brane. We will further comment on it later.

In (\ref{finalac}) the gauge coupling constant is 1. However, as shown
in \cite{bbn2}, in the path integral there is a non--trivial dependence on the
original gauge coupling $g$ which is due to the integration over the zero modes.
For our previous rescaling of the various fields by powers of $g$ involves, in 
particular, a rescaling of both the gauge and ghost diagonal degrees of freedom.
When defining the path integral we have to take this fact into account, which 
amounts to rescaling it by an overall factor for any given instanton. This 
factor is a power of $g$, the exponent being the number of zero modes for each 
rescaled field with the appropriate sign. It would seem therefore that we have 
to count the number of ghost and gauge zero modes. However this would lead us 
to a wrong result for the reason explained below.

\subsection{Summing over line bundles}

Eq. (\ref{finalac}) does not tell the whole story.  In fact in the previous 
subsection we have dropped the diagonal connection $\hat A$. Reintroducing
now this connection amounts to replacing ${\bf a}$ with ${\bf A+a}$,
where ${\bf A}$ is a non--trivial self--dual or anti--self--dual connection.
Since self--dual and anti--self--dual instantons lead to the same coverings,
when selecting a definite interpolating surface $\Sigma$ (to represent a
given scattering process) we have to allow for (i.e. to sum over) all
the ic--instanton solutions that contain such a surface as a covering, both
self--dual and anti--self--dual, and with all the possible non--trivial
connections ${\bf A}$. These are line--bundle connections (it is useful to
clarify that the fluctuation ${\bf a}$ is a 1--form: added to a line
bundle connection it supplies another connection; in the treatment of the
previous subsection it was supposed to be added to the 0 connection, i.e.
to represent a connection in the trivial line bundle over ${\Sigma}$;
in turn the fluctuations ${\bf x}^i$ as well as all the other fluctuating fields
are section of trivial line bundles). In conclusion we have to sum over
all line bundles on $\Sigma$ and integrate over all the connections in each
line bundle.

There is another way one can view the same problem:
we have to admit on $\Sigma$ any line bundle whose direct image under $\pi$ 
coincides with the initial vector bundle $E$ on ${\cal X}$. The construction
is, roughly speaking, as follows. In a covering with $N$ sheets, any line
bundle $L$ generates an $N$--component `vector' on ${\cal X}$: its components
are just 
the $N$ lines that lie over the same point of ${\cal X}$. Therefore to any line
bundle over $\Sigma$ there corresponds a vector bundle $E$ over ${\cal X}$.
The Chern classes $c_1(E),c_2(E)$ are connected to the Chern class of $L$
via the Grothendieck--Riemann--Roch theorem, but, in the case of a noncompact
manifold like ${\cal X}$, these constraints may become irrelevant.

A clarification is in order concerning $c_1(E)$ ($c_2(E)$ is trivial, therefore
$c_2(E)$ does not need a comment). A non--trivial first Chern class on a brane
world--volume is usually interpreted as the signal of the presence of a membrane. 
According to our interpretation membranes are not present in the theory, and
$c_1(E)$ is a pure geometrical feature of the base manifold, which must be 
considered on the same footing as the complex structure and the like. It is
only if we sum over all non--trivial $c_1(E)$'s on the base that we are allowed
to sum over all the line bundles on the covering.

The correspondence we have just outlined is described in more detail and with
more appropriate language in Appendix. Summarizing, the spirit of our approach 
implies that we have to allow for anything in $\Sigma$ can be lifted from
${\cal X}$, or, equivalently,  for anything in $\Sigma$ can be projected 
down to something that lives in ${\cal X}$.  Therefore, on $\Sigma$, we have to 
allow for all possible non--trivial line bundles.  The path integral 
must include the sum over such line bundles over $\Sigma$, as well as the path 
integration over all the connections on such line bundles.

A path integration with sum over all line bundles in a Maxwell theory 
has already been carried out in \cite{witten}, see also \cite{RW}, and we
follow this calculation. For the sake of clarity we partially reproduce it here.

The trick consists in passing to a dual formulation by introducing 
auxiliary fields and enlarging the gauge symmetry. Given a connection ${\bf A}$
on a line bundle ${\cal L}$, one first introduces the auxiliary field 
${\bf G}_{\mu\nu}$, and requires the 
theory to be invariant under an extended gauge symmetry whose local version
is
\be
{\bf A} \to {\bf A} + {\bf \Omega}, \quad\quad {\bf G} \to{\bf G} + d{\bf \Omega}
\label{gauge}
\ee
where ${\bf \Omega}$ is a local one--form. In addition (global version) one requires
that ${\bf G}$ be defined up to closed two--forms. This is tantamount to asking
that the integrals of ${\bf G}$ over two--cycles be defined up to integers.
Then, if ${\bf F}$ is the curvature of ${\bf A}$, one defines the combination
${\mathfrak F}_{\mu\nu}= {\bf F}_{\mu\nu} - {\bf G}_{\mu\nu}$. 
${\mathfrak F}$ is clearly invariant under the generalized gauge transformation
(\ref{gauge}) because ${\bf F}$ integrated over a two--cycle gives an integer. 

Now one considers the series of dual line bundles $\tilde {\cal L}$ with 
dual connection ${\bf V}_\mu$ and curvature ${\bf W}_{\mu\nu}$, and writes 
the action
\be
I = \frac {1}{2} \int_\Sigma d^4\xi \sqrt h \left(\frac{i}{4\pi}
\epsilon^{\mu\nu\lambda\rho} {\bf W}_{\mu\nu} {\bf G}_{\lambda\rho}
+ {\mathfrak F}^+_{\mu\nu} {\mathfrak F}^{+\mu\nu}+
{\mathfrak F}^-_{\mu\nu} {\mathfrak F}^{-\mu\nu}
\right)\label{I}
\ee
where + and -- denotes the self--dual and anti--self--dual part of a two form,
respectively. There are two alternatives. On the one hand, integrating over 
${\bf V}$ one obtains that $d{\bf G} =0$ 
and ${\bf G}$ has integral periods, which allows us to set ${\bf G}=0$,
in view of (\ref{gauge}); in this way we get back the original lagrangian
for ${\bf A}$. On the other hand, using (\ref{gauge}) one can simply set 
${\bf A}=0$, and end up with the dual formulation, where the basic fields
are the connection ${\bf V}$ and the two--form ${\bf G}$.

\subsection{Counting zero modes}

The dual formulation has the virtue of transforming the discrete summation over
line bundles into an integration over continuous fields. Now we can see that
the relevant zero modes are those of the one--forms ${\bf V}$ together
with the corresponding ghosts (the dual of ${\bf c}$), and the two forms
${\bf G}$. Looking at (\ref{I}) and at the definition of ${\mathfrak F}$ one 
sees that ${\bf V}$ rescales inversely with respect to ${\bf A}$, while
${\bf G}$ rescales in the same way. Therefore the zero modes of ${\bf G}$ 
and the ghosts of ${\bf V}$ will contribute with the same sign, while 
the zero modes of ${\bf V}$ 
will contribute with the opposite sign to the overall factor in front of the 
path integral. We are now ready to compute the latter.

Let us recall that our surface $\Sigma$ is a two--dimensional complex variety
with $n$ punctures (see section 4). If it were a compact surface we would say
that there are $2b_1$ zero modes of ${\bf V}$, two zero modes of the ghosts and
$b_2$ zero modes of ${\bf G}$, where $b_1,b_2$ are Betti numbers of $\Sigma$.
In conclusion we would get an overall factor $g^{2b_1-b_2-2}$. Now since
$2-2b_1 +b_2$ is the Euler characteristics of a compact 4D manifold, and since
a puncture takes away one unit of Euler characteristics (as can be seen 
for example by triangulating the manifold), we are led to the conclusion that
the overall factor in the presence of $n$ puncture is $g^{-\chi}$ where 
$\chi=2-2b_1+b_2 -n$. This is the correct result and there are several ways
one can convince oneself of it. The easiest one is probably by use of a doubling
construction, as in \cite{bbn2}. Let us first make more precise the concept of
puncture. The open surface $\Sigma$ has boundaries $B_i$ with $i=1,\ldots,n$,
which are 3--manifolds. For each $B_i$ let us consider the cone $C_i$, which
is obtained from the `cylinder' $B_i\times I$, where $I$ is a finite interval, 
by `squeezing' to a point one of the boundaries of the cylinder. We can attach
the boundaries of these cones to the corresponding boundaries $B_i$ of 
$\Sigma$
and obtain a compact surface $\Sigma_c$. Since each $B_i$ has Euler 
characteristic 1, using additivity of the latter, we get 
$\chi\equiv\chi_\Sigma=\chi_c -n$, where $\chi_c$ is the Euler characteristic 
of the compact surface,
i.e. $\chi_c=2-2b_1+b_2$. Now let us turn to the double of $\Sigma$:
one constructs a complex surface $\hat\Sigma$ endowed with an antianalytic 
involution (locally this is $z\to \bar z$). Roughly speaking the double is made 
of two copies of $\Sigma$ attached by the boundaries to form a compact surface: 
each boundary of one copy 
is attached to the corresponding boundary of the other copy. Denoting  
by a hat the quantities relevant to the double and using again additivity of 
the Euler characteristic, we have 
$$
\hat\chi= 2 -2\hat b_1 + \hat b_2= 2\chi= 4-4b_1+2b_2 -2n\0
$$
Now the number $2 -2 \hat b_1 +\hat b_2$ is the total alternating sum of zero
modes on the double. Since it can be expressed via the Gauss--Bonnet theorem as 
an integral over $\hat\Sigma$, we expect that the same integral over $\Sigma$
would yield half of it, i.e. $\chi$, thanks to the symmetry implied by the 
anti--involution. We expect therefore that the total number of zero modes
on $\Sigma$ with the appropriate sign be given by $\chi$, which is the result
anticipated above.
 
Now one can see that for all $\Sigma$'s considered in this paper as branched 
coverings of 
${\cal X}$, $\chi \geq 0$, therefore the sum over ic--instantons gives rise 
to a series
of non--negative powers \footnote{In MST the corresponding series is in terms
of $g^{\chi}$; however Riemann surfaces with at least two punctures have 
negative $\chi$} of $1/g$. This suggests that we interpret $g_3=1/g$ as the 
D3--brane perturbative interaction coupling, analogous to the string coupling
of \cite{bbn2}.

What we have shown so far is not enough to draw definite and unambiguous 
conclusions, however we have seen that in the strong coupling limit of 4D SYM 
theory there is room to describe scattering processes of D3--branes. Let us 
discuss a few general aspects of these processes. Ic--instantons become
four--manifolds with boundaries consisting of three--manifolds. These 
geometrical configurations lend themselves to an interpretation in terms of
three--brane scattering. In turn this interpretation fits very well in the path 
integral formalism, since it gives rise to a perturbative series in $1/g$. 
It remains for us to specify what are the amplitudes involved in these 
scattering of branes. Like in the case of string scattering, we will not
really mean scattering of full branes but rather scattering of particle
states which represent brane excitations. Although we do not know the spectrum 
of states of a D3--brane, in the case at hand we know plenty of
such states: all the fields ${\bf x}^i, {\bf a}_\mu$ as well as the fermions
which appear in (\ref{finalac}), together with their derivatives and products
are eligible to create such states. It is clear how to proceed: whenever
we want to represent scattering of 3--branes excitations with given
incoming and outgoing states, we have to insert
in the path integral the appropriate fields and evaluate the corresponding 
amplitudes. Of course this is not the end of the story, since one should then
sum over all the instantons that interpolate between the same initial and
final states, which means a sum over the appropriate instanton topologies and, 
at fixed topology, an integral over the appropriate moduli space. 

The states we have just mentioned are local states, i.e. they should be
associated to points of $\Sigma$, not to boundary 3--manifolds $B_i$. Suitable
4--manifolds are obtained by attaching cones $C_i$ to $B_i$, as explained above,
and smoothing out the result. Alternatively, we can imagine diffeomorphisms
that deform the boundaries to points, and restrict ourselves to such 
configurations. 

To end this section let us remark another difference with MST. While in MST
the Euler characteristics entirely determines the instanton topology, that is 
not so in the present case. In fact since $\chi =2- 2b_1 +b_2 - n$, there may be
and in fact there are manifolds with different $b_1$ and $b_2$ but the same
$\chi$. Therefore each term in the perturbative expansion in $1/g$ consist
in general of a sum over different topologies. This sum is potentially infinite.
However, as long as $N$ is finite, the number of topologies which is possible
to realize as branched coverings will be finite. Therefore $N$ may be considered
as a regulator for  these sums. In this regard there is an interesting
possibility: it is possible to introduce
a parameter that allows us to discriminate among the various terms of a sum
corresponding to a given $\chi$. This is $\theta$, the angle in front 
of the topological {\it theta term}, which can be introduced in the theory
in the usual way (see \cite{witten} for an analogous context). We will not 
do it here.

\section{Discussion}

In the course of the paper we have set aside a few problems which we
would like now to comment on. The first question we 
want to address is that of the interpretation of the scattering theory we 
have digged out in the previous sections. We have found several indications
that it is a scattering theory of D3--branes. Let us further justify this
claim in the light of Matrix theory.

\subsection{Connection with Matrix theory}

At least in the case ${\cal X}= {\mathbb R}\times {\mathbb T}^3$, the 
theory (\ref{eSYM}) can be easily derived from Matrix Theory, 
\cite{BFSS,bil,banks,BiS,WTIV}, via compactification on a dual 3-torus,
 \cite{WT}. Here we recall what is essential to make this connection, 
 following in particular \cite{FHRS} (see also \cite{sei,GS}). 
 
Matrix Theory hinges on the idea that a system of $N$ D0--branes infinitely 
boosted along a fixed direction, say the 11--th, describes the essential
features of M theory. Each D0--brane has the 11-th component of the momentum 
$p_{11}\sim 1/R_{11}$ far larger that the transverse components (where $R_{11}$
is interpreted as a large compactification radius for M theory). It is expected
that the Matrix Theory description of M theory becomes more and more
faithful as $N$ becomes larger and larger. Matrix Theory is represented
by supersymmetric quantum mechanics of matrices (SYM theory in 0+1 dimensions
with gauge group $U(N)$ and 16 supercharges).
Compactification of M--theory on a circle of radius, say, $R_9$ is expected to 
lead to IIA theory in the $R_9\to 0$ limit. In Matrix Theory the corresponding
operation consists in
compactifying the base manifold of SYM theory on the dual radius, so that 
one ends up with 1+1 dimensional SYM theory with $U(N)$ gauge group and 
${\cal N}=(8,8)$ supersymmetry, i.e. MST. That this leads to type IIA theory
in the strong YM coupling limit is by now a well--known result, which has been
recalled in the introduction. 

The next step is to compactify the base manifold of SYM theory along some
additional dimensions. Let us denote by $\tilde R_i$ the field
theory compactification radii and by $R_i$ the corresponding M--theory radii.
They are related by, see \cite{FHRS},
\be
\tilde R_i = \frac {\ell_{11}^3}{R_i R_{11}},\quad\quad i=9,8,...\label{dual}
\ee
where $\ell_{11}$ is the 11--th dimensional length scale.
For example, if one compactifies on a two--torus ($i=9,8$) and takes the limit
$R_9,R_8\to 0$, one can convince oneself that one series of massless states is
produced, which is interpreted as a new dimension that opens up. This
new dimension plus the one implicit in the large $N$ limit lead us back to 
10 dimensions in a type IIB framework. If, instead, we compactify on a 
three--torus ($i=9,8,7$) and  take the limit $R_9,R_8,R_7\to 0$ we can see that
three new dimensions open up: in this case dimensions and context are those of
M--theory. The latter however is not the limit we are interested in in this 
paper. We will rather consider the limit 
\be
R_7,R_8\to 0,\quad\quad R_9\to \infty.\label{limit}
\ee
It is easy to see that in this case only one series of massless states is 
produced, i.e. only one new dimension opens up (instead of three). Naturally
we have to add the decompactification related to $R_9$ being very large and the 
new dimension implicit in the large $N$ limit. Therefore the context of
(\ref{limit}) is that of a 10 dimensional type IIB theory. 
That it describes D3--branes can be seen by starting from Matrix Theory,
which is theory of D0--branes, compactifying along the 7,8,9--th directions
and t--dualizing the three circles. The D0--branes become D3--branes wrapped
around the three--torus. Finally one takes the limit (\ref{limit}) or the 
corresponding in the dual variables according to (\ref{dual}).

The YM theory we obtain is exactly (\ref{mSYM}). In fact the dependence on
the compactification radii can be entirely collected in the dimensionless
coupling constant
\be
g^2= \frac {\ell_{11}^3}{R_7R_8R_9 }.\label{coupling}
\ee
The action can be brought to the form (\ref{mSYM}) via a sequence of
rescalings.

In conclusion, the connection with Matrix Theory tells us that the scattering 
theory that looms through the previous sections, if confirmed by further analysis,
can be interpreted as a scattering theory of D3--branes in type IIB theory in 10D.

\subsection{Other questions}

In the previous subsection we have considered the case of 
$\mathbb R\times \mathbb {\mathbb T}^3$. 
Toroidal compactifications are the most well--known cases of 
compactifications of Matrix theory, \cite{WTIV}. Our point of view about Matrix 
Theory, however, is that its content is revealed and the information stored in 
it can be retrieved by considering all possible compactifications. This means 
that we should analyze other compactifications beside $\mathbb R\times \mathbb
{\mathbb T}^3$ and the ensuing ic--instantons in order to capture the full 
content of Matrix Theory. On the other hand, when studying a given D3--brane 
scattering process, we saw that ic--instantons with 
the same in and out configurations can come from different base manifolds.
This creates a potential problem: which is the right base manifold? 
One possible answer suggested by the previous considerations is
that we should perhaps sum over all instantons that interpolate 
between the relevant initial and final configurations, regardless of what 
base manifolds these instantons are originated from. 

One final comment concerns the comparison with the scattering of 
macroscopic D--branes mediated by open strings stuck on them, which is the way
scattering of macroscopic D--branes has been described in the literature up to 
now, \cite{pol}. This is an open problem, however the following remark might 
be helpful. 
In our approach a string mediated interaction would imply manifolds of 
real dimensions 
two being exchanged among the interacting strings, instead of manifolds of 
complex dimensions 
two, as in our case. We are clearly in the presence of a limiting (singular) 
case of the scattering described in this paper. This can be rephrased by 
saying that string--mediated D3--brane scattering amplitudes may be a limiting
case of the general scheme presented here: they may become the leading 
contributions under particular kinematical conditions.

\section*{Appendix. Mathematical description of the lifting.}

To really describe the process we called ``lifting'', it is convenient to
start from the opposite, i.e. how to push down a line bundle. Given a covering
$\pi: \tilde M \to M$, one has an easy way to get a line  bundle on $\tilde M$
from one on $M$, i.e. the pullback $\pi^*$; to ``push forward'' a line
bundle on $\tilde M$ we have to resort to the machinery of sheaves, in a way
that may, however, be easily translated in simple terms. 

Let ${\cal L}$ the line bundle on $\tilde M$, and denote with the same symbol
the sheaf of its holomorphic sections. Now we define the 
{\it direct image} sheaf $\pi_* {\cal L}$ on $M$ through the formula
\beq
\pi_* {\cal L}\ (U) = {\cal L} (\pi^{-1}(U)),
\eeq
where $U$ are open sets on the base $M$;
this, as we will discuss in a moment, is the sheaf of sections of a vector
bundle. The correspondence is this: the holomorphic sections of the vector 
bundle over $U \subset M$ are given by the holomorphic sections of the line
bundle over the open set $\pi^{-1}(U) \subset \tilde M$. \\
If $U$ is a disc that does not intersect the branch 
locus, $\pi^{-1}(U)$ consists of $N$ (the
order of the covering) distinct discs. The sections of ${\cal L}$ over these 
$N$ discs are simply $N$-uples of functions; these are interpreted as the
local sections of the vector bundle on the base, which therefore has rank
$N$.  
On a neighborhood of a branch point, the situation is different, since there
are less than $N$ discs. So it would seem that the rank changes, and that the
sheaf we defined does not correspond to a vector bundle. Let us analyze more
closely what happens in a situation of total branching, with a map from a disc
$\tilde U$
with coordinate $z$ to a disc $U$ with coordinate $w$, branched $k$ times: 
$z \mapsto z^k =w$. Any function on $\tilde U$ can be written, by the
Weierstrass preparation theorem, as $s(z) = s_1(z^k) + z s_2 (z^k) + \ldots
+ z^{k-1} s_k(z^k)$; the $s_i$ are functions of $w$, so any function on $U_z$
gives $k$ functions on $U$. This shows that the rank is constant: the local
sections of our direct image sheaf over a neighborhood of every point are $N$
holomorphic functions.

Consider now a section $y$ of the trivial line bundle over $\tilde
M$ (remember it is non compact, in our case; on a compact manifold, we would
have to take a non trivial line bundle and slightly modify the whole
construction, making another line bundle appear also on the base).
Multiplication by it gives a map $y: H^0(\pi^{-1}(U), {\cal L}) \to
 H^0(\pi^{-1}(U), {\cal L})$ and hence, by definition of $\pi_*{\cal L} \equiv
E$, a map that we call $X: H^0(U, E) \to  H^0(U, E)$. In a neighborhood of a
non branching point, a basis for the space of 
our local sections is given by local sections
$s_i$ of ${\cal L}$ over each of the discs $U_i$, such that ${s_i}_{|_{U_j}}$
for $i \neq j$. On this basis, $X$ acts as follows: $X \, s_i = y_{|_{U_i}}
s_i$. This means that the $y_{|_{U_i}}$ are the eigenvalues of $X$, and $s_i$
the eigenvectors. On neighborhoods of the branch points, we would have to
choose a different basis of sections, as described above; however, by
continuity it is still true that $y$ is given by the spectrum of $X$. 

Suppose now that we have a connection on ${\cal L}$: this means, for each 
tangent vector $v$, an endomorphism $D_v$ of the space of sections of ${\cal
L}$. To obtain a connection on $E$, consider again the situation locally
around a non branching point $p \in M$; given a tangent vector 
$v_p$, $(D_{v_p}\, s)_i$ is the $N$-uple having as components $D_{v_i}\ s_i$,
where $v_i$ are the $N$ counterimages of $v_p$. This means that we do the same
analysis we did for $y$ for each component of $a$, where $d +a$ is a local
expression for the connection. \\
In a disc containing a branch point, the analysis is different due to the
different basis; we don't describe explicitly the computations here, but let
us briefly sketch the result. Pushing down the connection gives, in general,
a connection on the base with poles; this gives relations between the Chern
classes of ${\cal L}$ and of $E$, which agree with the results one can find by
the Grothendieck-Riemann-Roch theorem. On our non compact manifolds, however,
most of these conditions are uneffective.

Now that we have described what happens going downstairs, we can try to invert
the process. In our situation, we simply have a section $X$ of a bundle
$End(E)$ over the base manifold $M$, a connection over this bundle, and
fluctuations. 
We have to reconstruct the covering manifold, the line bundle ${\cal L}$, 
the connection over it, and the map $y$. \\
The manifold is described by the spectrum of $X$: in $M\times
\cc$, it is given by the equation $\det (X - y) = 0$, and $y$ is a well-defined
function over it (it is one of the coordinates of the ambient space $M \times
\cc$). To obtain the line bundle, consider the pullback $\pi^* X$, section of
$\pi^* End(E)$, which is a rank $N$ vector bundle over $\tilde M$. The 
eigenspace 
$\ker(X -Y) \subset \pi^* E$ 
defines, completing by continuity also over the branching 
points, a line bundle ${\cal L}$; this is clearly the inverse of the
construction we gave above, since we observed that the $s_i$ were indeed 
eigenvectors of $X$ -- we just put on $M$ the eigenspace corresponding to
its eigenvalue. \\
What remains to be lifted is the connection, and all the fluctuations of the
fields. The process is similar, and we describe it for the connection.
Take the connection on $M$: it is a connection on $End(E)$, but we may
consider it as a connection on $E$ (it is a matter of choosing the
representation on which it acts). Pull it back to $\tilde M$: it defines a
connection on $\pi^* E$. Since we know that it commutes with $X$, it preserves
eigenvalues, and so it defines a connection on the line bundle. This is done
with the eigenvalues of $A$, and so what we described is nothing but the
formalization of the process described in the text. On the branches, the
process is different, as we mentioned above; but its analysis is beyond our
scope now, since we know the connection outside branches by what we described,
and over the branches we know that the lifting gives the right number of delta
functions to make the relations between the Chern classes of ${\cal L}$ and
$E$ match.

\vskip 1cm 
{\bf Acknowledgements.} We would like to thank F.Nesti for 
collaboration in the preliminary stage of this work and R.Dijkgraaf, 
P.A.Marchetti, P.Pasti,
C.Reina, A.Zampa and, in particular, D.Sorokin for helpful discussions.
L.B. would like to thank the Departamento de Fisica, Universidade Federal do
Espirito Santo, Vitoria, for the kind hospitality extended to him during
the early stage of this work.
This work was partially supported by EC TMR Programme,
grant FMRX-CT96-0012, and by the Italian MURST for the program
``Fisica Teorica delle Interazioni Fondamentali''.

\end{document}